# Tensile Deformation and Failure of Thin Films of Aging Laponite Suspension[§]

Asima Shaukat, Yogesh M. Joshi[*] and Ashutosh Sharma[*]

Department of Chemical Engineering

Indian Institute of Technology Kanpur, Kanpur 208016 INDIA

* To whom correspondence should be addressed.

For YMJ: E-mail address: joshi@iitk.ac.in.

For AS: E-mail address: ashutos@iitk.ac.in

**Abstract**

In this paper we study deformation, failure and breakage of visco-elastic thin films of aging laponite suspension under tensile deformation field. Aqueous suspension of laponite is known to undergo waiting time dependent evolution of its micro-structure, also known as aging, which is accompanied by an increase in the elastic modulus and relaxation time. In the velocity controlled tensile deformation experiments, we observed that the dependence of force and dissipated energy on velocity and initial thickness of the film is intermediate to a Newtonian fluid and a yield stress fluid. For a fixed waiting time, strain at break and dissipated energy increased with velocity, but decreased with initial thickness. With increase in age, strain at break and dissipated energy showed a decrease suggesting enhanced brittle behavior with increase in waiting time, which may be caused by restricted relaxation modes due to aging. In a force controlled mode, decrease in strain at failure at higher age also suggested enhanced brittleness with increase in waiting time. Remarkably, the constant force tensile deformation data up to the point of failure showed experimental time- aging time superposition that gave an independent estimation of relaxation time and elastic modulus dependence on age.

# I. Introduction:

Understanding deformation behavior of thin films under tensile deformation field is important for many applications such as pressure sensitive adhesives,[1] bioadhesives,[2] lubricant films,[3] hip joints,[4] cosmetic and pharmaceutical creams,[5] corneal tear film,[6] etc. The materials that constitute such films are often visco-elastic soft materials with complex microstructure caused by physical or chemical crosslinking and/or by presence of metastable phases. High viscosity and elasticity of these materials impose significant constraints on the translational mobility of primary structural entity that comprises the soft materials, forcing it to explore only a part of the phase space available to it.[7, 8] Rheological study of such systems is always challenging, since such materials often show yield stress and thixotropic characteristics.[9] In addition the broken ergodicity of these materials imparts aging which leads to strong history dependence. In this paper, we study the force response and dissipated energy in the tensile deformation leading to failure and breakage of thin films of aqueous suspension of laponite. Laponite suspension is a model thixotropic visco-elastic material showing strong aging behavior, which allows us to also investigate the influence of material visco-elasticity on its tensile deformation field. The behavior of such materials under tensile loading has not been investigated previously.

The class of soft materials that show ergodicity breaking due to jamming of the primary entity are represented as soft glassy materials.[10] Although the thermal motion is not adequate to attain equilibrium, the jammed entities undergo microscopic dynamics of structural rearrangement in their arrested state and explore those configurations that take them to a lower energy state.[11] This phenomenon is commonly known as aging and is accompanied by significant increase in viscosity and elasticity of the material as a function of time. Prominent characteristic feature common in these materials is extremely slow relaxation dynamics with dominant mode growing with the age of the system.[12] Application of the sufficiently strong deformation field imparts diffusion of the arrested entities out of their cages causing local yielding events which leads to decrease in relaxation time and its dependence on age.[10] This phenomenon is commonly known as rejuvenation. Therefore, aging and rejuvenation affects the visco-elasticity of these materials, which makes these materials to demonstrate complex rheological behavior. Aqueous suspension of smectite clay laponite is considered as a model soft glassy material that shows all the above mentioned characteristic features,[8] and hence is a subject of intense investigation over past decade. Various prominent studies on this system involve optical characterization to explore its relaxation dynamics[13-20] and phase behavior[15, 21-24] while various rheological tools, which essentially investigate shear flow behavior, have been employed to study aging under deformation field and thixotropic behavior.[25-30]

In many applications, thin viscous and visco-elastic layers undergo tensile deformation. Thus, deformation behavior of simple liquids such as silicon oil, oil paints, glycerol, castor oil, olive oil etc. in a Hele-Shaw cell or in similar geometries under tension has been studied.[31-34] Finger formation during the cohesive or adhesive failure of cross-linked visco-elastic liquid and solid polydimethylsiloxane (PDMS) has also been extensively investigated.[35-43] Principle focus of such studies is to study either the variation of force required or strain induced under variety of loading conditions or the morphology of the fingering patterns observed during the adhesive or cohesive failure. If the bonding between the plates and the fluid layer is weak, contact breaks at the boundary leading to debonding or adhesive failure.[35, 44-48] On the other hand, cohesive failure occurs if the contact area of the fluid that holds two plates together decreases causing failure in the bulk of the material.[32-35, 44-48] For the liquid-like films, at the higher timescales of plate separation, pressure gradients drive air into the fluid causing fingering patterns due to Saffman-Taylor instability.[32-35, 48, 49] For elastic solid-like films, adhesion failure also occurs by fingering, but their spacing is governed by the minimization of the stored elastic strain energy.[35-42] Thus, most of the studies concern rheologically simpler systems such as Newtonian,[32-35, 44, 48, 50, 51] linear viscoelastic,[34, 35, 44-47] and yield stress fluids[48, 49] and their corresponding fingering instability patterns under velocity and force controlled deformation fields. In Newtonian fluids, the relationship between pulling force $F$ and constant velocity $V$ can be obtained by solving Navier Stokes equations under tensile deformation and is given by,[48, 51]

$$F = 3\pi\mu R_i^4 d_i^2 V / 2d^5 ,\qquad(1)$$

where $R_i$ and $d_i$ are initial radius and thickness of the sample respectively, $\mu$ is viscosity and $d$ is the distance between two plates. Derks *et al.*[48] extended the analysis to yield stress fluid described by Herschel-Bulkley model in the limit of small deformation rates and obtained an expressing for pulling force given by,

$$F = 2\pi\mu R_i^3 d_i^{3/2} \sigma_y / 3 d^{5/2} ,\qquad(2)$$

where $\sigma_y$ is the yield stress. They observed an excellent agreement of this expression with the experimental behavior of systems having varying yield stress. Integration of above expression for force over a plate separation yields energy dissipated during tensile deformation. For a yield stress fluid, Derks *et al.*[48] observed it to be a linear function of yield stress while independent of the initial separation. The nano-clay suspension we study also shows yield stress. However unlike the system used by Derks *et al.*,[48] this nonlinear suspension demonstrates strong time and deformation field dependent rheological behavior, including aging of its rheological properties.[26, 27]

In this work, we investigate the force and energy response of aqueous visco-elastic suspension of laponite to applied tensile deformation field by varying the material visco-elasticity (aging time) and initial thickness under two distinct modes of deformation: constant rate of deformation (velocity) and constant normal force. A time-aging time superposition of the suspension data is employed to give an independent estimate of the dependence of the modulus and relaxation time on aging time. The process of suspension-film breakup is characterized by the strain and critical force required at the initiation of failure and the energy dissipated for a complete cohesive failure of the film. The effects of visco-elasticity (aging time) and strength of deformation field on the mode of material failure, such as brittle or viscous, are also addressed. Results are compared and contrasted with the model systems of the Newtonian and yield stress fluids.

**II. Material and Experimental procedure:**

Laponite is a synthetic hectorite clay and belongs to the structural family known as the 2:1 phyllosilicates.[52] Laponite is commonly used in the chemical and the food industry to control rheological properties of the end product. Laponite is composed of disc shaped particles with a diameter 25 nm and a layer thickness 1 nm.[53] The chemical formula for laponite is $Na_{+0.7}[(Si_8Mg_{5.5}Li_{0.3})O_{20}(OH)_4]_{-0.7}$. The surface of laponite has a permanent negative charge while the edge is less negative in the basic pH medium while positive in the acidic pH medium.[52] At pH 10 the negative charge on the surface leads to overall repulsion among the laponite particles,[15] although the attractive interactions between the edge and the surface can not be ruled out.[54] Soon after mixing laponite with water, system undergoes ergodicity breaking.[55]

Laponite RD used in this study was procured from Southern Clay Products, Inc. Laponite powder was dried for 4 hours at 120°C before mixing with ultra pure water at pH 10 under vigorous stirring. The basic environment is necessary to provide chemical stability to the suspension.[56] The pH 10 was maintained by the addition of NaOH. The suspension having 3.5 weight % of laponite was stirred vigorously by Ultra turrax T25 for about 1 hour and left undisturbed for a period of about 3 months in a sealed polypropylene bottle. The rheological and tensile tests were carried out using the parallel plate (50 mm diameter) geometry of Anton Paar MCR 501 rheometer. Before placing the suspension on the lower plate of the rheometer, the sample was shear melted (rejuvenated) by passing it through a syringe needle several times. Shear melting is needed to achieve a uniform initial state before beginning the experiment and validation of the same by passing through a syringe needle was confirmed by reproducibility of the initial state. Subsequently, the upper plate was lowered until the gap between both the plates attained a predetermined value. After keeping the sample unperturbed for a desired waiting time, that was measured since predetermined gap was set, the upper plate was pulled in a direction normal to the plates by two modes namely a constant velocity and a constant force. In MCR 501 rheometer, the normal force transducer is located at the bottom plate and has a least count of $10^{-2}$ N. In a force controlled mode, the normal force on the bottom plate is controlled by the movement of the top plate. In order to avoid aging during the experiment, the experimental time was kept significantly smaller than the waiting time (age) of the sample. We have carried out the waiting time dependent experiments only up to waiting time of 15 min, as beyond 20 min sample started to dry at the rim of the film. In all the experiments temperature was maintained at 25 °C.

**III. Results and Discussion**

Aqueous suspension of laponite is known to undergo aging in which its elastic modulus increases with the waiting time (age).[27] Figure 1 shows an evolution of elastic and viscous modulus with waiting time in a small amplitude oscillatory experiment ($\gamma_0$=1%, and frequency = 0.1 Hz). It can be seen that $G'$ increases with waiting time and the dependence can be approximated by a power law given by, $G' \sim t_w^{0.15}$. This behavior is generally addressed as aging and is due to rearrangement of the arrested particles within the cage to attain a lower energy state.[11] If we approximate the material response as a time dependent single mode Maxwell model,[26, 27] dominating relaxation time of the material can be related to elastic and viscous modulus as: $\tau = G'/\omega G''$.[27] We have also plotted relaxation time in figure 1, which obeys a stronger power law dependence ($\tau \sim t_w^{0.27}$) on waiting time compared to elastic modulus. Although the relaxation time is estimated by considering only a single mode of the Maxwell model, it gives a good first estimate.[27]

As mentioned before, in this work we have employed two modes: a constant velocity of the top plate and a constant pulling force to study tensile deformation behavior. A typical response to applied constant velocity is shown in figure 2, wherein normal force showed a rapid increase in the beginning followed by a slow decline as a function of time. The point of initiation of failure and the point of complete breakage are also shown in the figure. In addition, we have plotted the distance ($d$) between the plates as a function of time on the same figure. Except for a very short period after starting the experiment (< 1 s), the rheometer maintained a constant velocity of the upper plate. It was observed that, due to separation of plates, cross-sectional area of the suspension reduced to a fractal like branched patterns. After the failure, the fractal pattern that connected both the plates formed fibrils and eventually underwent breakage. We discuss various characteristic features of these fractal like branched patterns (see figure 11) later in this section. For all the experiments carried out in this paper, the breakage occurred by cohesive failure in the bulk of the material.

Figure 3 shows strain at break ($\Gamma_b$) and strain at failure ($\Gamma_f$) plotted against applied constant velocity. These experiments were carried out on the films having thickness around 100 µm and at waiting time of 15 min. Strains at failure for low velocities are not shown because at the time of failure, the top plate had not attained a constant velocity. It can be seen that the strain at break increased with increase in velocity. Inset in figure 4 shows variation of force as a function of movement of the top plate ($d - d_i$, where $d$ is distance between two plates and $d_i$ is initial gap), for various values of velocities. It can be seen that force at failure also increases with velocity of top plate. Area under these curves represents energy dissipated in the tensile deformation experiments. As shown in figure 4, dissipated energy per unit initial area of the film increased with increase in velocity of the top plate and the dependence can be represented by a power law given by: $E \sim V^{0.2}$. As discussed in the introduction, for a Newtonian fluid, dissipated energy varies linearly with velocity, while yield stress fluid with Herschel Bulkley constitutive relation shows dissipated energy to be independent of velocity. The present system showed a much weaker response compared to what is expected for a Newtonian liquid.

An initial gap between the two plates is also an important variable that affects the severity of deformation field. We studied the effect of the same at constant velocity of top plate (50 µm/s) and constant waiting time (15 min). The corresponding variation of strain at failure ($\Gamma_f$) and breakage ($\Gamma_b$) is plotted in figure 5. It can be seen that both the strains, $\Gamma_f$ and $\Gamma_b$, decreased with increase in the initial gap. The corresponding energy dissipated per unit initial area, plotted in figure 6, also showed decrease with increased initial film thickness with a power law dependence $E \sim d_i^{-0.36}$. This dependence is also significantly weaker than that for a Newtonian fluid ($E \sim d_i^{-2}$) and is closer to a yield stress fluid for which the dissipated energy is independent of the initial plate spacing. It is apparent from figures 3 to 6 that the variation of strain at failure and break and that of dissipated energy is similar for increase in initial gap and for decrease in top plate velocity. One of the ways to assess the role of visco-elasticity is to define a Deborah number, $De = (\tau V/d_i)$, where the ratio, $d_i/V$ represents the timescale of initial deformation and $\tau$ is the relaxation time of the suspension described in figure 1. A large value of $De$ (> 1) denotes a dominantly elastic response, whereas $De$ < 1 indicates viscous behavior. Figure 7 shows strain at break and energy dissipated as a function of Deborah number. Strain at break, as well as energy dissipated, show increase with increase in Deborah number (or with decrease in timescale of deformation). It is known from the mechanical energy balance of a flowing fluid that the dissipated energy is given by: $\sum_i \sum_j \sigma_{ij} \nabla_j u_i$,[57] where $\sigma_{ij}$ is a stress tensor, $\nabla_j$ is a vector differential operator and $u_i$ is a velocity vector. Hence, viscous dissipation increases with a decrease in the timescale of deformation or increase in $De$, which is also equivalent to an increase in $\nabla_j u_i$. Therefore, these results suggest that with increase in time scale of deformation, contribution to the dissipated energy primarily comes from viscous effects. In addition, at lower timescales of deformation, a higher strain can be supported before failure and breakage because of increased importance of dissipation and decreased elastic storage of the energy.

As shown in figure 7, strain at break and dissipated energy do not depend solely on the initial timescale of deformation ($d_i/V$), but individually on $V$ and $d_i$.

We believe that three factors are responsible for such behavior. Firstly, the energy dissipated for a viscous fluid scales as $V/d_i^2$.[48] On the other hand, the yield stress material is independent of initial time scale of deformation, $(d_i/V)$.[48] These trends suggest that dependence of normal force on velocity and initial thickness is different than just $(d_i/V)$ and hence shows different responses as shown in figure 7. The second possible reason is the continuous increase in the time scale of deformation as experiment progresses due to increase in the distance between the plates, which causes continuous decrease in $De$. Therefore, various values of initial thickness and velocity, though leading to same ratio of $(d_i/V)$ may affect the tensile deformation field differently causing changes in the strain and dissipated energy. In essence, $De$ decreases with time, whereas the values given in figure 7 are only the initial values at the start of the experiments. Third important reason is the thixotropic character of the present system. Since the viscoelastic behavior of aqueous suspension of laponite is highly sensitive to the deformation field,[26] partial rejuvenation of the same reduces its viscosity as the deformation progresses. This behavior, in addition to the reasons discussed above, makes the overall behavior of the suspension highly complex, and as shown in figure 7, does not depend solely on the initial timescale of deformation, $(d_i/V)$ or initial $De$.

We further carried out the tensile deformation at different waiting times (ages) of the sample for constant velocity and initial suspension thickness. In figure 1, we have observed that elastic modulus of aqueous laponite suspension increased with waiting time (age). It is generally observed that for this class of materials, yield stress is also proportional to its elastic modulus, and thus, yield stress should also increase with time.[48] The effect of aging time ($t_w$) was studied for the suspension ages that are significantly larger than the time taken in the tensile test, $t_w \gg t$, so that further aging during the experiment could be neglected. Figure 7 shows that strain to break and dissipated energy both decline with $De$ or the relaxation time, which is proportional to aging time of the suspension. The discussion below considers the effect of aging in detail.

Figure 8 shows variations of strain at failure and strain at break as a function of age. Both the strains decreased with the waiting time. As expected, the experiment associated with zero waiting time shows deviation from the general trend because aging occurred during the test and the viscoelastic properties of suspension changed over the duration of experiment. For all the other experiments reported, we have maintained $t_w \gg t$. Figure 9 shows variation of force as a function of displacement of the top plate. The inset in this figure shows a double logarithmic plot of the same data. For a Newtonian fluid and a Herschel-Bulkley model, decay in force is expected to follow power law relations, $F \sim (d-d_i)^{-5}$ and $F \sim (d-d_i)^{-5/2}$, respectively. However, the laponite suspension showed further weaker dependence than the yield stress fluid. Engineering stress at failure and energy dissipated per unit initial area are plotted as a function of waiting time in figure 10. It can be seen that tensile stress at failure (or force at failure) closely followed same dependence on waiting time as that of elastic modulus. On the other hand, unlike Herschel-Bulkley prediction,[48] wherein energy dissipated increases with yield stress or elastic modulus, energy dissipated in the clay suspension decreased with increased waiting time and hence with increase in the elastic modulus. This result is also expected from figure 8, which shows that increase in waiting time or elastic modulus caused decreased strain at break. In addition, it is apparent from the nature of force displacement curve shown in figure 9 that the response of material became more brittle with increase in waiting time. It can be seen that for the force-deformation curves at higher waiting times of 10 min and 15 min, force dropped significantly after its maximum value. The severity of drop is more for a higher waiting time sample. We have represented this by a gray arrow on figure 9. We believe that the observed behavior is due to corresponding increase in elastic modulus and relaxation time with waiting time. Enhancing elastic modulus limits the dissipation while increasingly slower relaxation process causes progressively more limited modes of energy dissipation, leading to the brittle failure in the material. We have plotted the corresponding strain to break and dissipated energy as a function of Deborah number in figure 7. It can be seen that the waiting time dependent data shows opposite trend compared to the velocity and initial gap dependent data obtained at constant age. While former experimental data showed enhanced brittle failure due to increased elasticity the latter data showed pronounced viscous dissipation with increase in velocity and/or decrease in the initial thickness. This observation further strengthens our claim that, due to various reasons discussed above, the tensile deformation of the thin films of the aging suspension leading to its breakage does not solely depend on the Deborah number and is independently influenced by the relaxation time, elasticity, velocity and initial thickness of the sample.

In the tensile deformation experiments on visco-elastic materials, failure occurs by fingering or fractal patterns.[32, 34, 35, 47, 58, 59] We also observed identical fractal or branched patterns on both the plates upon complete separation of the plates. Figure 11 shows photographs of the top plate immediately after the complete cohesive failure in the waiting time dependent experiments. It can be seen that the contact lines, where the breakage occurred, became more branched with increase in waiting time. We have also listed perimeter of the suspension- air interface per unit plate area as well as number of branches of the same. It can be seen that both the properties that characterize intensity of the branched structure, increased with the waiting time. We also observed similar enhancement in branching with increase in the top plate velocity and with decrease in the initial gap. The effect of initial gap and top plate velocity on branching of the contact line is in accordance with the Derk et al.'s[48] observations ascribed to the Saffman-Taylor instability.[49] Derk et al.[48] observed that intensity of fingers do not have any direct correlation with dissipated energy. Interestingly, we observed that for various experiments performed at constant waiting time by varying velocity and initial thickness, increase in the intensity of branches was accompanied by increase in dissipated energy. On the other hand, for the experiments carried out as a function of waiting time, with fixed velocity and initial suspension thickness, increase in the intensity of branches showed decrease in the dissipated energy. Thus, there seems to be two different factors at work determining the branching propensity. This observation is in accordance with different trends in strain at break and dissipated energy as a function of Deborah number. Increasing the suspension elasticity or the speed of crack or finger propagation, both induce more frequently bifurcating and shorter cracks. The effect of higher failure speed on denser finger patterns is also seen in simple liquids.[31, 33, 35, 48, 59]

We next discuss tensile deformation of the aging suspension of laponite under constant force. Figure 12 shows a typical variation of distance between the plates under application of constant force. In MCR 501 rheometer, stress (at the bottom plate) is maintained by controlling the velocity of the top plate. As shown in the figure, material starts to fail at a certain critical strain and the rheometer cannot maintain a constant force thereafter. The time to failure and strain at failure mark the point at which force starts decreasing. It can be seen from figures 13 and 14 that both the time to failure and strain at failure decreased with increase in applied force and initial thickness. For a Newtonian fluid, relationship between the time of separation (which is approximately equal to time to failure) for the constant pulling force is given by:[34, 50, 60] $t_s = 3\pi\mu R_i^4 / 2Fd_i^2$ . As shown in figures 13 and 14, similar to that observed for the constant velocity experiments, response of the laponite suspension is weaker compared to a Newtonian fluid.

Finally, we perform constant force experiments at various waiting times. Figure 15 shows time to failure and strain at failure as a function of waiting time. Interestingly, strain at failure decreases with the waiting time (we are disregarding the point associated with zero age, as $t \ll t_w$). Since force is constant, strain at failure is directly proportional to energy dissipated in causing failure in the sample, which also decreases with the waiting time. This further strengthens our observation that increase in elastic modulus induces brittleness in the system. Time to failure increases with increase in waiting time showing power law dependence, $t_f \sim t_w^{0.3}$ . This dependence is very close to the dependence of relaxation time on waiting time, thus suggesting a linear dependence of time to failure on relaxation time $\left(t_f \sim \tau\right)$. In figure 16, tensile strain is plotted against experimental time for three waiting times. It can be seen that the younger sample shows enhanced strain at the same experimental time ($t$). Dominant relaxation mode of soft glassy materials such as an aqueous suspension of laponite is known to show a power law dependence on waiting time.[12] The corresponding power law exponent, $\left(\mu = d\ln\tau/d\ln t_w\right)$, which represents the rate of increase of relaxation time, depends on the creep stress.[26, 61, 62] Since the relaxation processes usually intrinsically determine the rheological behavior of the material, the normalization of creep time by an appropriate characteristic timescale is expected to give superposition of the experimental data after appropriate vertical shifting is carried out. Figure 17 demonstrates that a horizontal and vertical shifting of tensile compliance vs. time data indeed shows a successful superposition. Tensile creep compliance is obtained by dividing strain in figure 16 by normal stress $(J(t) = \Gamma(t)/\sigma_N)$. We have also plotted horizontal and vertical shift factors as a function of aging time in the inset of figure 17. Considering time dependent single mode Maxwell model, Joshi and Reddy[26] suggested that in order to obtain a superposition, the vertical shift factor should scale as modulus, while the horizontal shift factor should scale as inverse of relaxation time. Remarkably, horizontal shift factor ($a^{-1} \sim \tau \sim t_w^{0.31}$) and vertical shift factor ($b \sim G' \sim t_w^{0.15}$) do indeed closely match respective dependences of relaxation time and elastic modulus on age obtained from the dynamic experiments shown in figure 1. This procedure of obtaining a universal behavior is known as creep time-aging time superposition and is generally used to obtain independent estimation of relaxation time and elastic modulus dependence on age.[63] Such procedure is often used in polymeric glasses to predict long time tensile strain from short time tests.[63, 64] Various groups have demonstrated validity of the same concept in shear flow of soft glassy materials. However, according to best of our knowledge this is the first report of validity of the same to the *tensile* deformation of soft glassy materials.

## IV. Conclusions

In this paper we have studied deformation, failure and breakage of thin films of aging suspension of laponite under tensile deformation field. We have applied tensile deformation using two modes, namely constant velocity of top plate and constant pulling force as a function of initial thickness of the sample and its age. Aqueous suspension of laponite shows time dependent evolution of visco-elastic properties, also known as aging, wherein elastic modulus and relaxation time increased with increase in waiting time. Under the tensile deformation field, stretching of the suspension beyond a critical strain lead to the initiation of failure in the form of finger/crack/fibril formation and eventually a complete cohesive failure in the bulk of the sample with fractal like features on the freshly formed surfaces. In a constant velocity mode, strain at failure and strain at break increased with increase in velocity, but decreased with increase in initial thickness of the sample. In addition, the energy dissipated in the tensile deformation leading to breakage of the film increased with increase in velocity of suspension-stretching and also with decrease in initial suspension-film thickness. However, the dependences for both trends were intermediate between what is expected for a Newtonian fluid and a yield stress fluid. We believe that in the velocity and initial gap dependent experiments, increased intensity of the deformation field induces more viscous dissipation in the material. Interestingly with increase in waiting time or age, energy dissipated and strain at failure and breakage showed a decrease, demonstrating enhanced elastic or brittle response at the higher age. We believe that this behavior is caused by increase in elasticity and relaxation time with the waiting time that restrict the modes of energy dissipation leading to rapid breakage. Overall, due to continuously changing intensity of the flow field and thixotropy of the material the tensile deformation behavior is observed to be not control by initial Deborah number.

In a constant pulling force mode, dependence of time to failure on magnitude of force and initial thickness showed a weaker response than what is expected for a Newtonian fluid. For the waiting time dependent experiments, decrease in strain at failure for older samples also indicated enhanced brittleness with age. Interestingly experimental data of tensile creep compliance normalized with modulus at different waiting times showed a superposition when plotted against experimental time normalized with relaxation time of the system. The validation of this time-aging time superposition by tensile deformation data demonstrated universal applicability of this procedure for soft glassy materials.

**Acknowledgement**: Financial support from Department of Science and Technology through IRHPA scheme is greatly acknowledged. YMJ also acknowledges partial support from Department of Atomic Energy, BRNS young scientist award scheme.

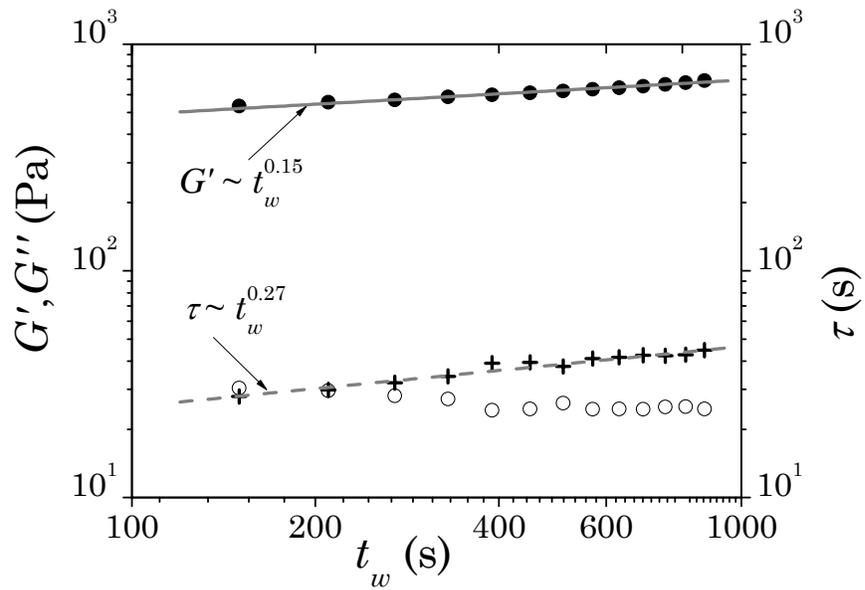

Figure 1. Evolution of elastic modulus ($G'$, filled circles), viscous modulus ($G''$, open circles) and relaxation time ($\tau$, +) as a function of waiting time ($f$ =0.1 Hz, $\gamma_0 = 1$ %). Thick gray line represents a power law fit to elastic modulus – waiting time data $\left(G' \sim t_w^{0.15}\right)$, while dashed gray line represents a power law fit to relaxation time – waiting time data $\left(\tau \sim t_w^{0.27}\right)$.

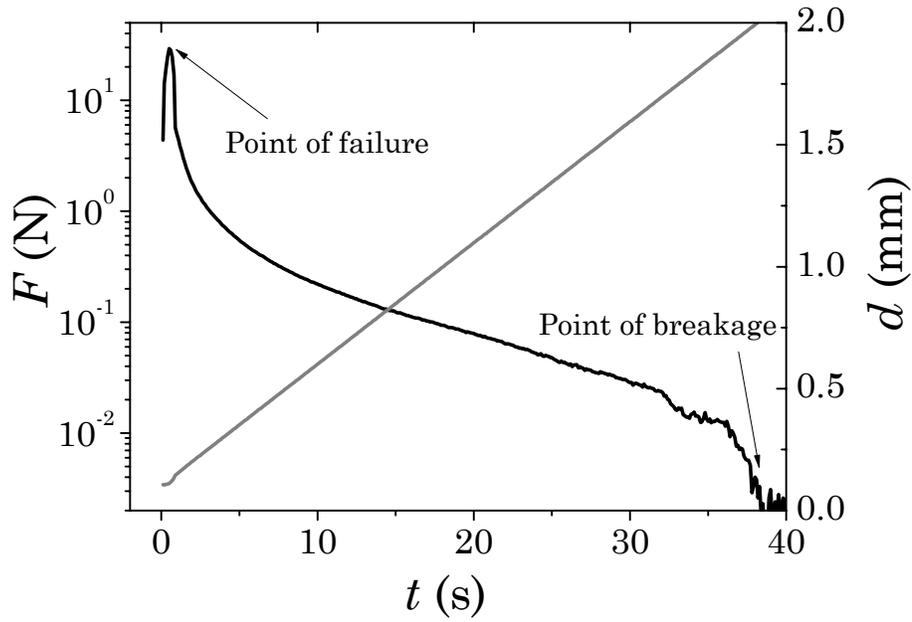

Figure 2. Variation of the normal force $F$ (thick black line) and distance between the plates (thick gray line) as a function of experimental time $t$ for $V=50$ μm/s, $t_w=15$ min and $d_i=107$ μm.

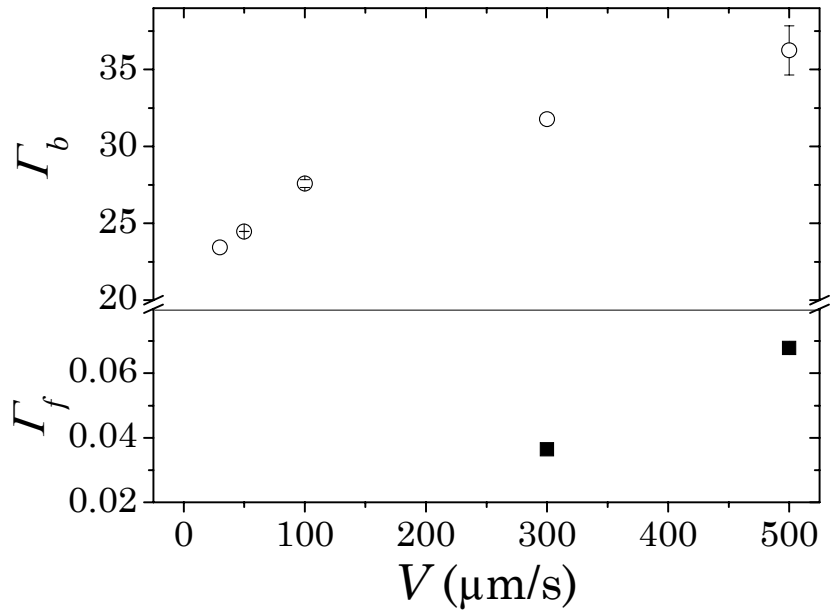

Figure 3. Strain at failure and break as a function of velocity of the top plate for $t_w=15$ min and $d_i=105\pm 2$ μm.

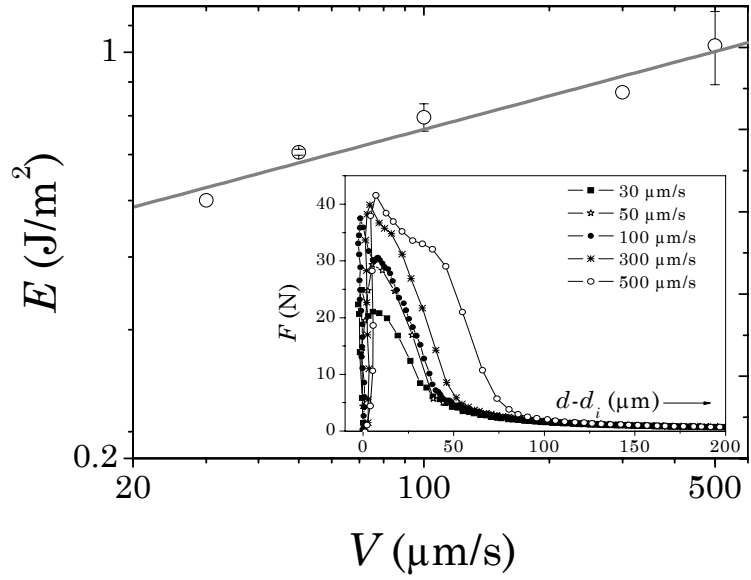

Figure 4. Dissipated energy as a function of velocity of the top plate for $t_w$=15 min and $d_i$=105±2 μm. Thick gray line shows a power law fit to the experimental data given by: $E \sim V^{0.2}$. In an inset force is plotted against displacement of the top plate for various velocities. Area under the curve gives energy dissipated in the tensile deformation.

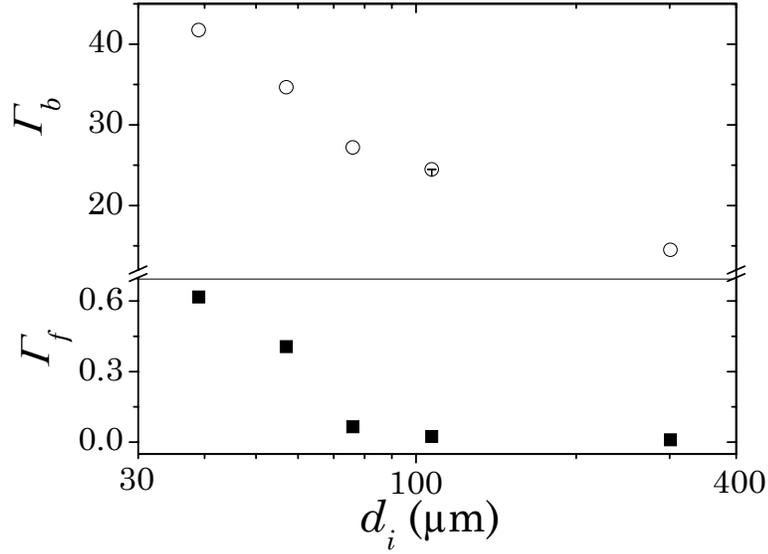

Figure 5. Strain at failure and break as a function of initial gap, $d_i$ for $t_w$ =15 min and $V$ =50 µm/s.

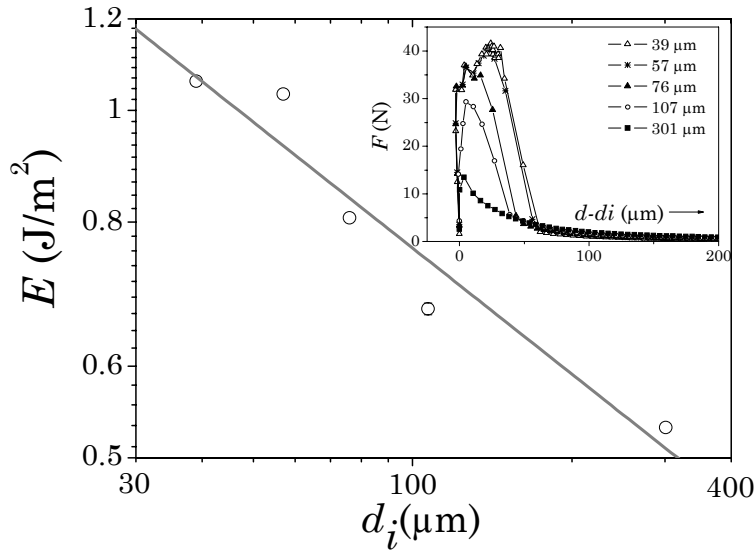

Figure 6. Dissipated energy as a function of initial gap, $d_i$ for $t_w$ =15 min and $V$ =50 µm/s. Thick gray line shows a power law fit to the experimental data given by: $E \sim d_i^{-0.36}$. In an inset force is plotted against displacement of the top plate for various initial gaps.

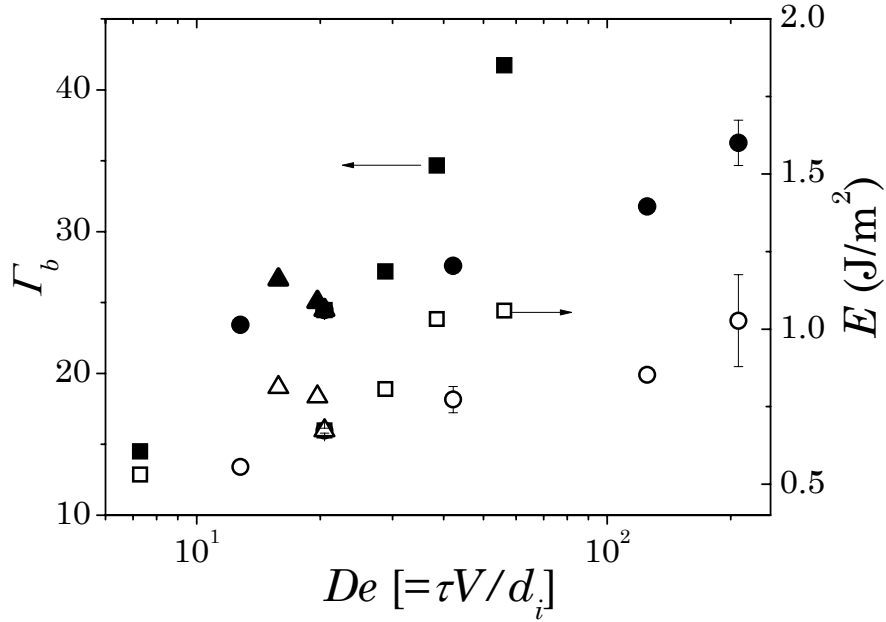

Figure 7. Strain at break (filled symbols) and energy dissipated (open symbols) in the tensile deformation leading to breakage of the film is plotted against Deborah number based on time scale of deformation for various velocity (circle), initial thickness (square) and waiting time (triangle) dependence experiments.

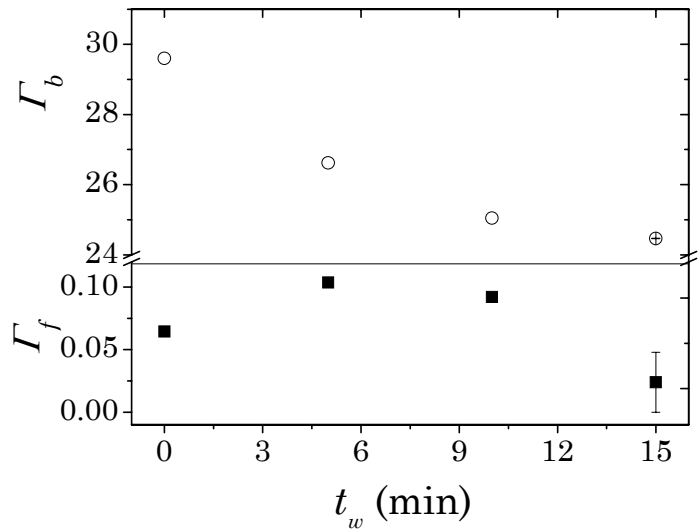

Figure 8. Strain at break as a function of waiting time for $d_i=106\pm 1$ μm and $V=50$ μm/s.

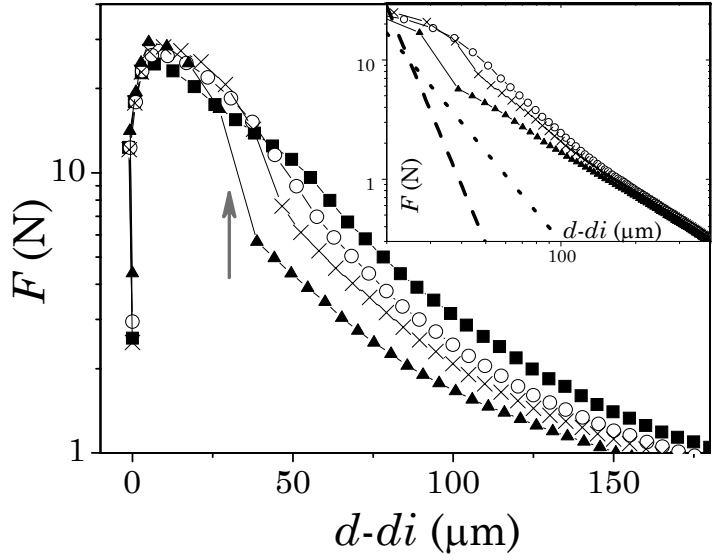

Figure 9. Force as a function of displacement of top plates for experiments carried out at various waiting times ($t_w$=0 min: filled squares, 5 min: open circles, 10 min: cross and 15 min: filled triangles) for $V$ =50 µm/s and $d_i$=106±1 µm. Gray arrow points a sudden decrease in force over a small displacement of top plate at higher waiting times. Inset shows same plot on a double logarithmic scale. Thick dashed and dotted lines in the inset represent a power law prediction for a Newtonian $\left[F \sim (d-d_i)^{-5}\right]$ and a yield stress fluid (Herschel-Bulkley) $\left[F \sim (d-d_i)^{-5/2}\right]$ respectively.

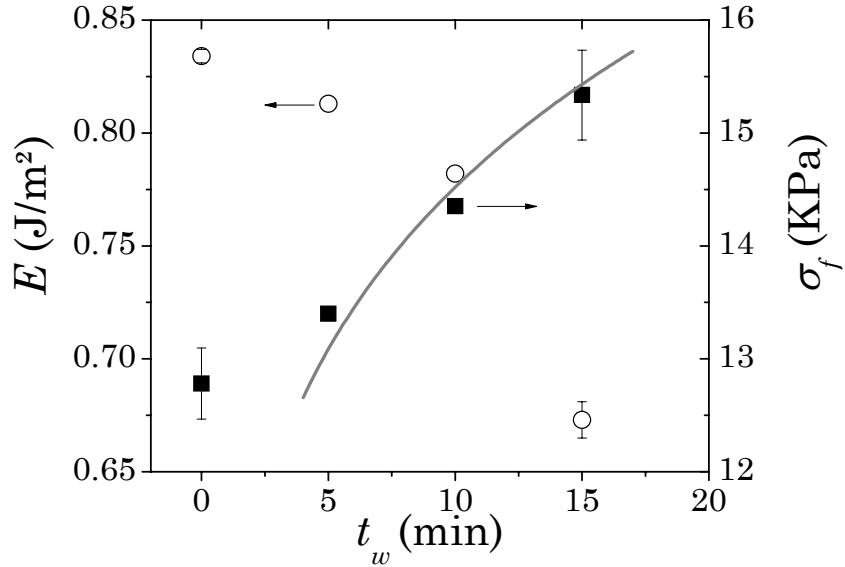

Figure 10. Energy dissipated (open circles) and stress at failure (filled squares) as a function of waiting time for $V = 50$ μm/s and $d_i = 106 \pm 1$ μm. Thick gray line represents a power law fit to the engineering stress at failure data representing $\sigma_f \sim t_w^{0.15}$.

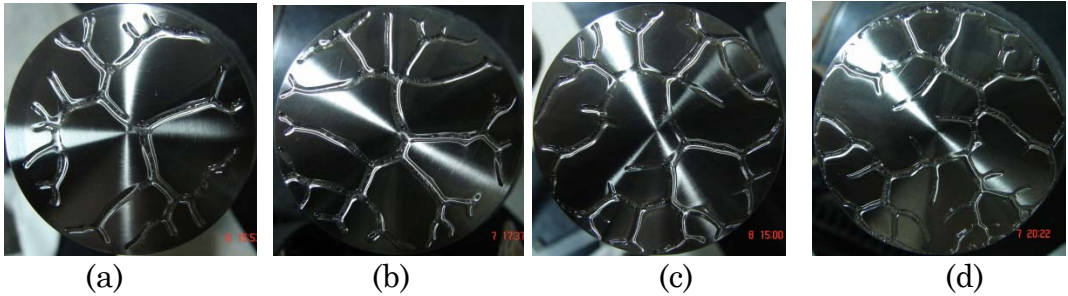

(a) (b) (c) (d)

Figure 11. Photo of the upper plate after breakage. (a) $t_w = 0$ min (Perimeter of the suspension- air interface per unit area = 0.2458 /mm, Number of branches=35), (b) $t_w = 5$ min (Perimeter of the suspension- air interface per unit area = 0.3215 /mm, Number of branches=47), and (c) $t_w = 10$ min (Perimeter of the suspension- air interface per unit area = 0.3492 /mm, Number of branches=54) and (d) $t_w = 15$ min (Perimeter of the suspension- air interface per unit area = 0.4034 /mm, Number of branches=63). All the experiments were carried out at $d_i = 106 \pm 1$ μm and $V = 50$ μm/s.

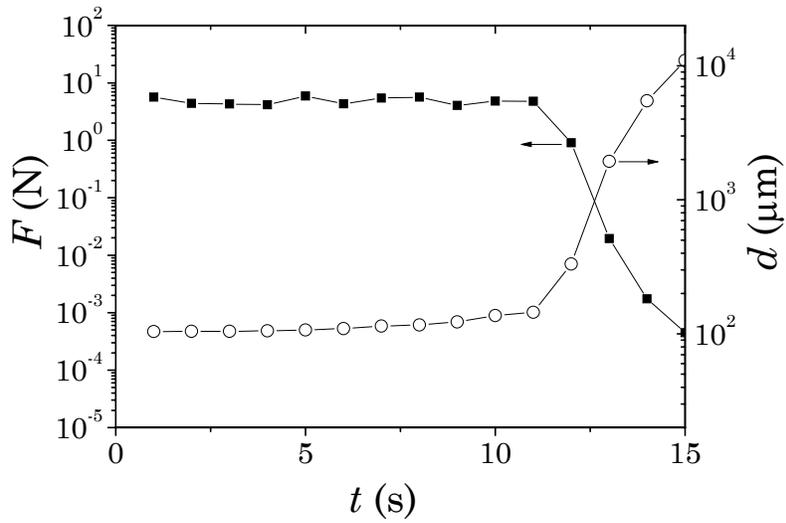

Figure 12. Pulling force measured at the bottom plate ($F=5$ N) and resultant variation of the gap between the plates as a function of experimental time $t$ for $t_w=15$ min, $d_i=100$ μm. Filled squares represent force while open circles represent gap.

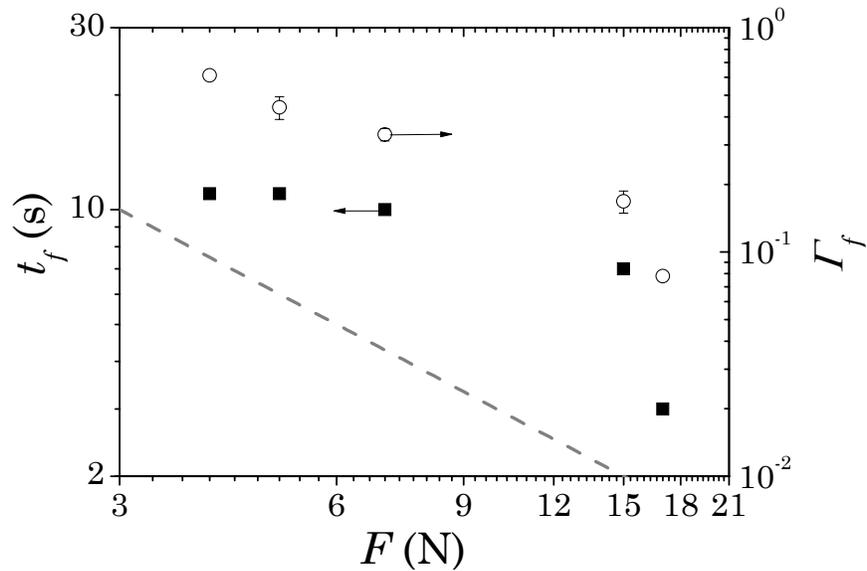

Figure 13. Time to failure (filled squares) and strain at failure (open circles) as a function of force applied to the top plate for $t_w=15$ min, $d_i=103\pm2$ μm. Thick gray line represents prediction for a Newtonian fluid given by $t_f \sim F^{-1}$.

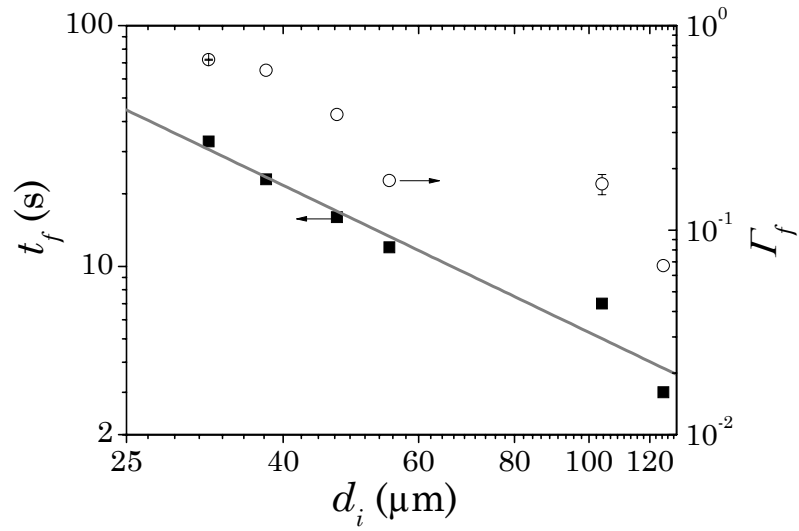

Figure 14. Time to failure (filled squares) and strain at failure (open circles) as a function of initial gap for $t_w$ =15 min and $F$ =15 N. Thick gray line shows a power law fit to time to failure data representing $t_f \sim d_i^{-1.5}$.

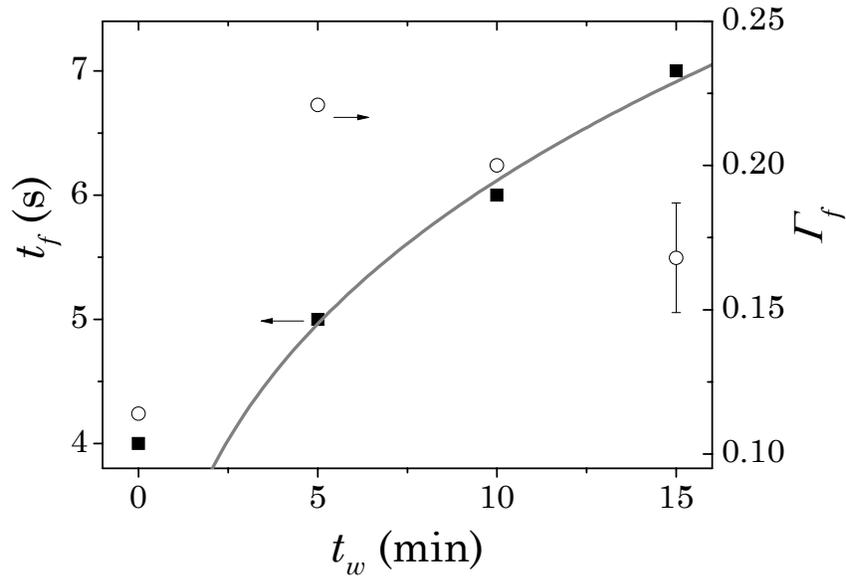

Figure 15. Time to failure (filled circles) and strain at failure (open circles) as a function of waiting time for $F=15$ N and $d_i=104\pm 1$ µm. Thick gray line represents a power law fit to time to failure dependence on waiting time given by: $t_f \sim t_w^{0.3}$.

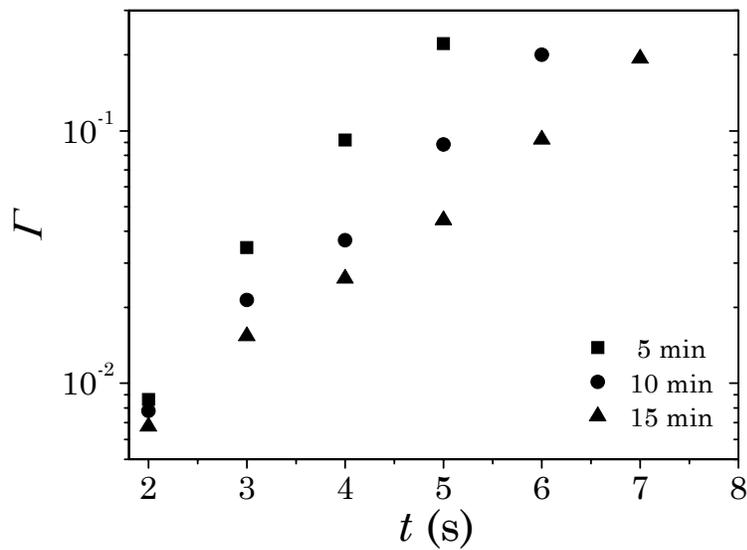

Figure 16. Tensile strain as a function of experimental time at constant force of 15 N and at different waiting times ($d_i=104\pm 1$ µm).

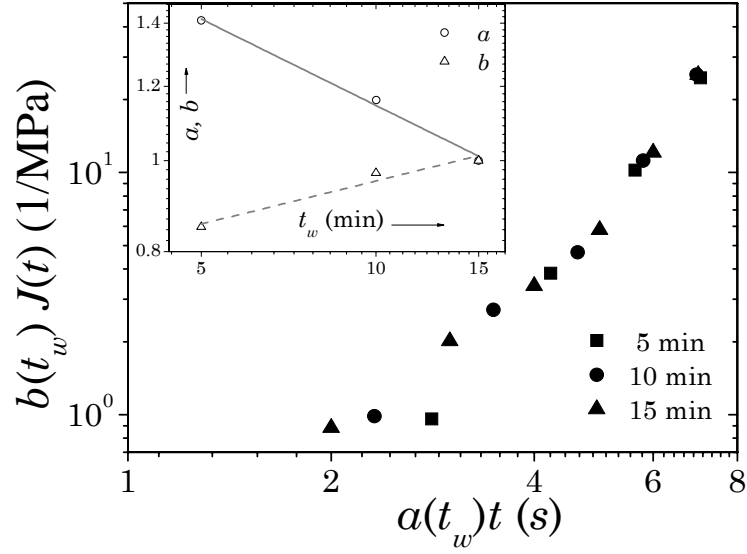

Figure 17. Superposition after carrying out horizontal and vertical shifting of experimental data shown in figure 15. Ordinate in figure 15 is divided by constant stress to yield compliance $J(t)$. Inset shows that horizontal and vertical shift factors show power law dependence on waiting time given by, $a \sim t_w^{-0.31}$ and $b \sim t_w^{0.15}$ respectively.